# Hemodynamic Markers: CFD-Based Prediction of Cerebral Aneurysm Rupture Risk

Reza Bozorgpour[1], Jacob R. Rammer[1]*

Department of Biomedical Engineering, University of Wisconsin-Milwaukee, Milwaukee, WI, USA

*Corresponding author
E-mail: jrrammer@uwm.edu

## Abstract

This study investigates the influence of aneurysm evolution on the hemodynamic characteristics within the sac region. Using computational fluid dynamics (CFD) methods, blood flow patterns through the parent vessel and aneurysm sac were analyzed to assess the impact of aneurysm evolution on wall shear stress (WSS), time-averaged wall shear stress (TAWSS), and the oscillatory shear index (OSI)—key indicators of rupture risk. Additionally, Relative Residence Time (RRT) and Endothelial Cell Activation Potential (ECAP) were examined to provide a more comprehensive understanding of the aneurysm's hemodynamic environment. Six distinct cerebral aneurysm (CA) models, all from individuals of the same gender, were selected to minimize gender bias in the analysis.

The results revealed that unruptured aneurysm cases exhibited higher WSS and TAWSS values, along with lower OSI and RRT values, consistent with stable flow conditions that promote endothelial health and vascular wall integrity. Conversely, ruptured aneurysm cases demonstrated lower WSS and TAWSS values, coupled with elevated OSI and RRT values, suggesting disturbed and oscillatory flow patterns that are commonly associated with aneurysm wall weakening. Furthermore, ECAP values were notably higher in ruptured cases, indicating increased endothelial activation driven by unstable hemodynamic conditions. Interestingly, regions with the highest OSI and RRT values often aligned with organized vortex centers, reinforcing the link between disturbed flow dynamics and aneurysm instability.

These findings emphasize the significance of combining multiple hemodynamic parameters to improve the assessment of aneurysm rupture risk. The inclusion of RRT and ECAP offers additional insight into the complex interplay between flow patterns and endothelial activation, providing a more robust framework for evaluating aneurysm stability and informing clinical decision-making.

## 1. Introduction

Cerebral aneurysms are weakened or abnormally bulging regions in the brain's blood vessels. These aneurysms may experience structural changes over time, which can significantly impact their hemodynamics, or the patterns of blood flow within them [1-4].

Hemodynamic variations significantly influence the development of cerebral aneurysms and can impact their rupture risk. Aneurysms typically begin as minor, localized dilations in blood vessels, and as they expand, both their structure and hemodynamic behavior may change. The aneurysm's geometry, including its shape, size, and orientation, plays a key role in determining blood flow characteristics. These hemodynamic changes are broadly classified into two categories: steady-state and transient flow alterations. Steady-state changes arise as the aneurysm enlarges, increasing the volume of blood inside it. This enlargement can alter flow velocity and wall shear stress, potentially weakening the aneurysm wall and heightening rupture risk [5-7].

Transient flow variations arise from dynamic changes in the surrounding hemodynamic environment. Factors like fluctuations in heart rate, shifts in blood pressure, and changes in blood flow volume can alter the internal flow behavior of the aneurysm. These variations may trigger intricate flow patterns such as vortex formation, flow recirculation, and regions of intensified wall shear stress. Over time, these disturbed flow dynamics can deteriorate the aneurysm wall, increasing the likelihood of rupture [8-10].

The dimensions of an aneurysm significantly impact its hemodynamic behavior and rupture potential. As an aneurysm grows, the forces exerted on its walls become more intense, placing greater strain on the vessel structure. This increase in wall tension elevates the likelihood of rupture. Furthermore, larger aneurysms are often associated with more intricate and unstable blood flow patterns, which can further heighten the risk [8, 11-13].

A clear understanding of how cerebral aneurysms evolve, and the resulting hemodynamic changes are essential for making [14-17]. Clinicians rely on advanced imaging techniques and computational tools to analyze aneurysm characteristics and assess blood flow dynamics. This valuable information supports the development of appropriate treatment plans, including surgical or endovascular interventions, to reduce the likelihood of rupture and improve patient prognosis [18, 19].

CFD plays a crucial role in analyzing hemodynamic factors associated with cerebral aneurysm development [20-25]. By simulating fluid flow through mathematical modeling, CFD helps researchers examine the intricate blood flow dynamics within cerebral vessels [26-28]. In aneurysm studies, CFD models are constructed using medical imaging data to create virtual representations of the affected vessels. These models enable the prediction of hemodynamic parameters, offering insights into how blood flow behaves inside the aneurysm and its surrounding vascular network.

WSS is a critical hemodynamic parameter that reflects the frictional force blood exerts on vessel walls. Elevated WSS has been linked to aneurysm progression and rupture risk [6, 29-33]. CFD simulations are instrumental in mapping WSS distribution within aneurysms, helping to pinpoint regions where elevated stress may signal potential rupture sites. Additionally, CFD provides detailed visualizations of complex flow behaviors, such as vortices, recirculating patterns, and unstable flow structures. These dynamic patterns can indicate areas of heightened wall stress or points vulnerable to rupture. Understanding these flow characteristics allows clinicians to assess aneurysm behavior more effectively and determines potential risks [34-36].

CFD is also valuable in assessing the effectiveness of various treatment strategies for cerebral aneurysms. By simulating blood flow both before and after interventions like stent placement or flow-diverting devices, CFD helps evaluate how these treatments influence hemodynamics. This information supports clinicians in determining the most suitable treatment approach for individual patients. However, it is important to recognize that CFD is a modeling technique with inherent limitations [37, 38]. The accuracy of CFD outcomes depends heavily on the quality of input data and model parameters, as the simulations often rely on assumptions and simplifications. As a result, CFD findings should be interpreted carefully and considered alongside clinical expertise and other diagnostic methods. Overall, CFD serves as a powerful tool for analyzing hemodynamic factors in cerebral aneurysm progression. It facilitates the estimation of wall shear stress, the visualization of complex flow patterns, and the evaluation of treatment outcomes. Incorporating CFD into clinical decision-making can improve the understanding of aneurysm behavior and contribute to better patient management strategies.

Despite numerous studies exploring aneurysm rupture risks across various sizes and locations, the relationship between rupture risk and aneurysm evolution remains insufficiently addressed in existing literature. This study aims to bridge that gap by conducting numerical simulations of blood flow in CA using a custom-developed Lattice Boltzmann Method (LBM) solver. Six CA models, comprising three ruptured and three stable aneurysms, are analyzed under identical conditions. The investigation focuses on examining key hemodynamic factors, specifically WSS, TAWSS and OSI, throughout the aneurysm's growth process.

## 2. Numerical Method

### 2.1. LBM

In LBM, fluid flow is modeled through the movement of hypothetical fluid particles. The distribution function, denoted as $f$, represents the particle number density within a specific particle group. The number of these groups is finite, typically denoted as $Q$. For simulating isothermal and incompressible flows in three-dimensional space, the D3Q19 model (where $Q = 19$) is commonly employed (Fig. 1). Instead of directly solving the Navier–Stokes and continuity equations, the LBM predicts fluid motion by calculating the time evolution of the discrete distribution function, $f_i$, corresponding to the $i\text{-}th$ particle group. The governing equation for this time evolution is known as the lattice Boltzmann equation.

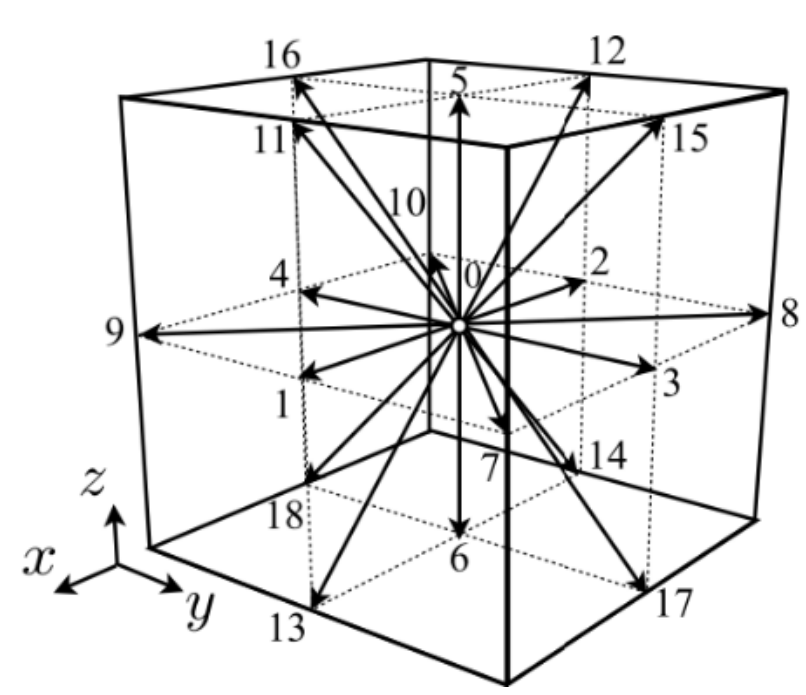

Fig. 1: D3Q19 discrete velocity model [39]

$$f_i(x + c_i\Delta t, t + \Delta t) = f_i(x, t) - \frac{f_i(x,t) - f_i^{eq}(x,t)}{\tau_{SRT}} \qquad (1)$$

In this context, $\tau_{SRT}$ represents relaxation time, while $x$ denotes the position vector. The time step size, $\Delta t$, is typically set to unity for simplification. The equilibrium distribution functions, $f_i^{eq}$, are defined by the following expression:

$$f_i^{eq} = \omega_i \rho [1 + 3c_i.u + \frac{9(c_i.u)^2}{2} - \frac{3u.u}{2}] \qquad (2)$$

In this formulation, $\omega$ represents the weighting function, defined as follows: $\omega_0 = 1/_3$, $\omega_1 = 1/_{18}$ for $i =$ 1-6, and $\omega_0 = 1/_{36}$ for $i =$ 7-18. The particle velocity is denoted by $c$. The grid spacing, $\Delta x$, is set to unity, ensuring that particles travel from one computational cell to its neighboring cell during the time step $\Delta t$. The density $\rho$ and the fluid velocity $u$ are calculated as moments of the distribution function $f_i$, expressed as:

$$\rho = \sum_{i=0}^{Q-1} f_i \qquad (3)$$

$$\rho u = \sum_{i=0}^{Q-1} c_i f_i \qquad (4)$$

The solution of the lattice Boltzmann equation involves two key steps. The first step is the collision step, which is performed at each computational cell. During this step, particle distribution functions within a cell interact and redistribute according to the specified collision model, typically based on the Bhatnagar-Gross-Krook (BGK) [40] approximation or the single-relaxation-time (SRT) model. This step is crucial for ensuring the correct exchange of momentum and energy among particles, ultimately influencing the fluid's macroscopic behavior.

$$\tilde{f}_i(x, t) = f_i(x, t) - \frac{f_i(x,t) - f_i^{eq}(x,t)}{\tau_{SRT}} \qquad (5)$$

$$f_i(x + c_i\Delta t, t + \Delta t) = \tilde{f}_i(x, t) \qquad (6)$$

The viscosity, $\mu$, is given by

$$\mu = \frac{\rho}{3}(\tau_{SRT} - \frac{1}{2}) \qquad (7)$$

## 2.2. LBM grid generation

In our approach to geometry processing and fluid simulation, we developed a specialized C++ code to manage these tasks effectively. The STL file, representing the triangulated surface of the vasculature, was obtained from MRA imaging [41-44]. Initially, the process involved reading the STL file and calculating the bounding box dimensions by determining the minimum and maximum coordinates of the triangular facets, ensuring the geometry was fully enclosed. The method began with a simplified cylindrical geometry to introduce the ray-casting concept, as shown in Fig. 2. While the image does not depict all points within the bounding box for clarity, imagine that points are uniformly distributed throughout the bounding box in practice. To illustrate the ray-casting method clearly, points A, B, and C are highlighted as representative points for classification. Using the ray-casting method, a ray is projected from each point in a chosen direction, and the number of triangle intersections is counted. If the count is odd (B), the point is identified as fluid; if even (A and C), it is marked as non-fluid. This approach efficiently distinguishes between internal and external points while maintaining visual clarity in the figure.

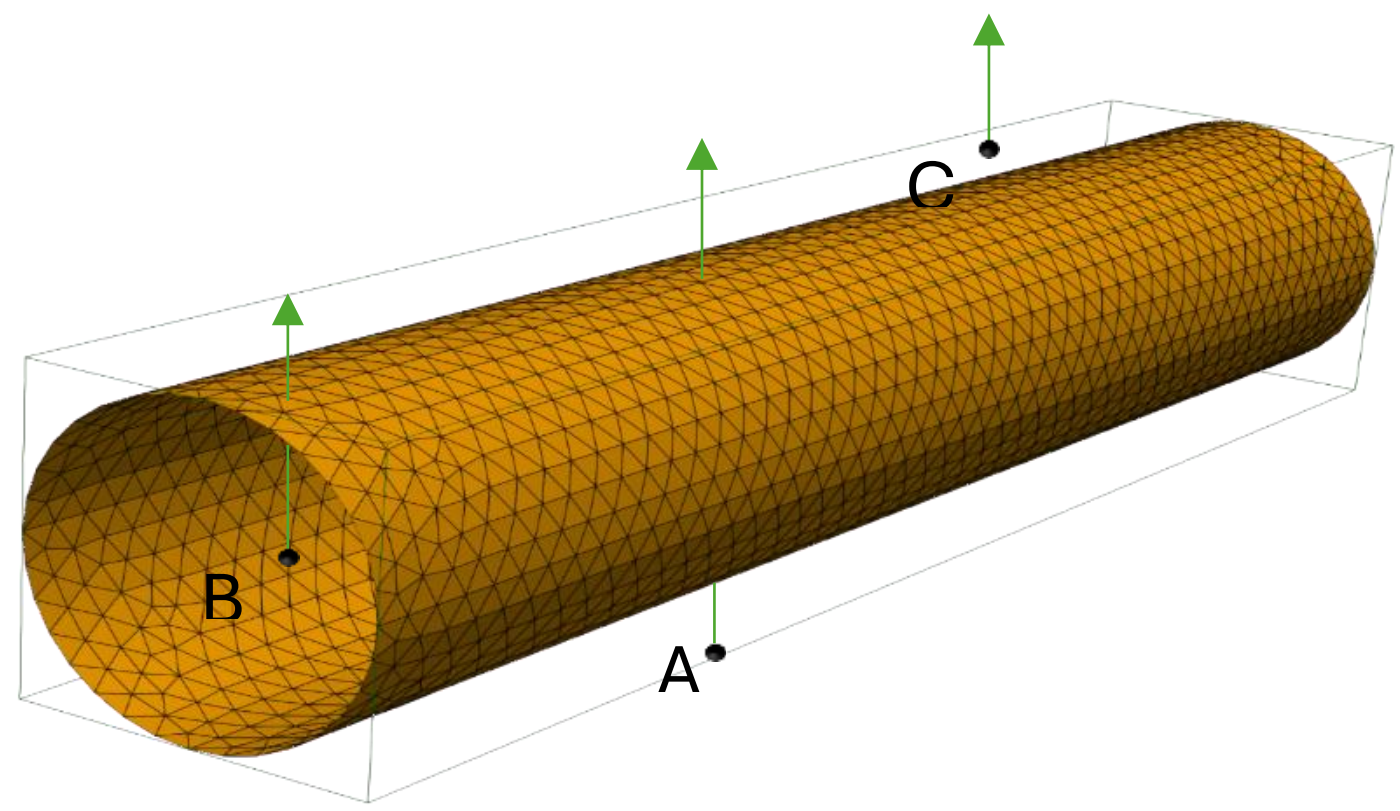

Fig. 2: Ray-casting method illustrated on a cylinder. Points A, B, and C represent uniformly distributed points in the bounding box. Rays are projected to classify points as fluid (odd intersections) or non-fluid (even intersections).

To improve the method's accuracy and efficiency, we advanced this approach by integrating MPI parallelization. Instead of generating a single large bounding box, we determined the coordinates of the global bounding box and distributed the computational workload across MPI processors. Each processor was assigned a portion of the bounding box and the triangulated vasculature surface, ensuring better workload distribution and scalability. This enhanced method is illustrated in Fig. 3, where the vascular geometry is enveloped by distributed lattice points. Additionally, ghost cells were incorporated along the processor boundaries to maintain consistency during boundary point identification, which classified fluid points adjacent to non-fluid points as boundary points. This refined approach ensured greater accuracy and computational efficiency, particularly for complex geometries like cerebral vasculature.

The simulations were performed on the University of Wisconsin-Milwaukee's high-performance computing cluster. This system consisted of eight AMD compute nodes, each equipped with two 64-core AMD EPYC Milan 7713 processors running at 2.0 GHz and 256 GB of RAM. With 128 cores per node and a total of 1024 cores across the system, the cluster provided substantial computational power for handling complex simulations efficiently.

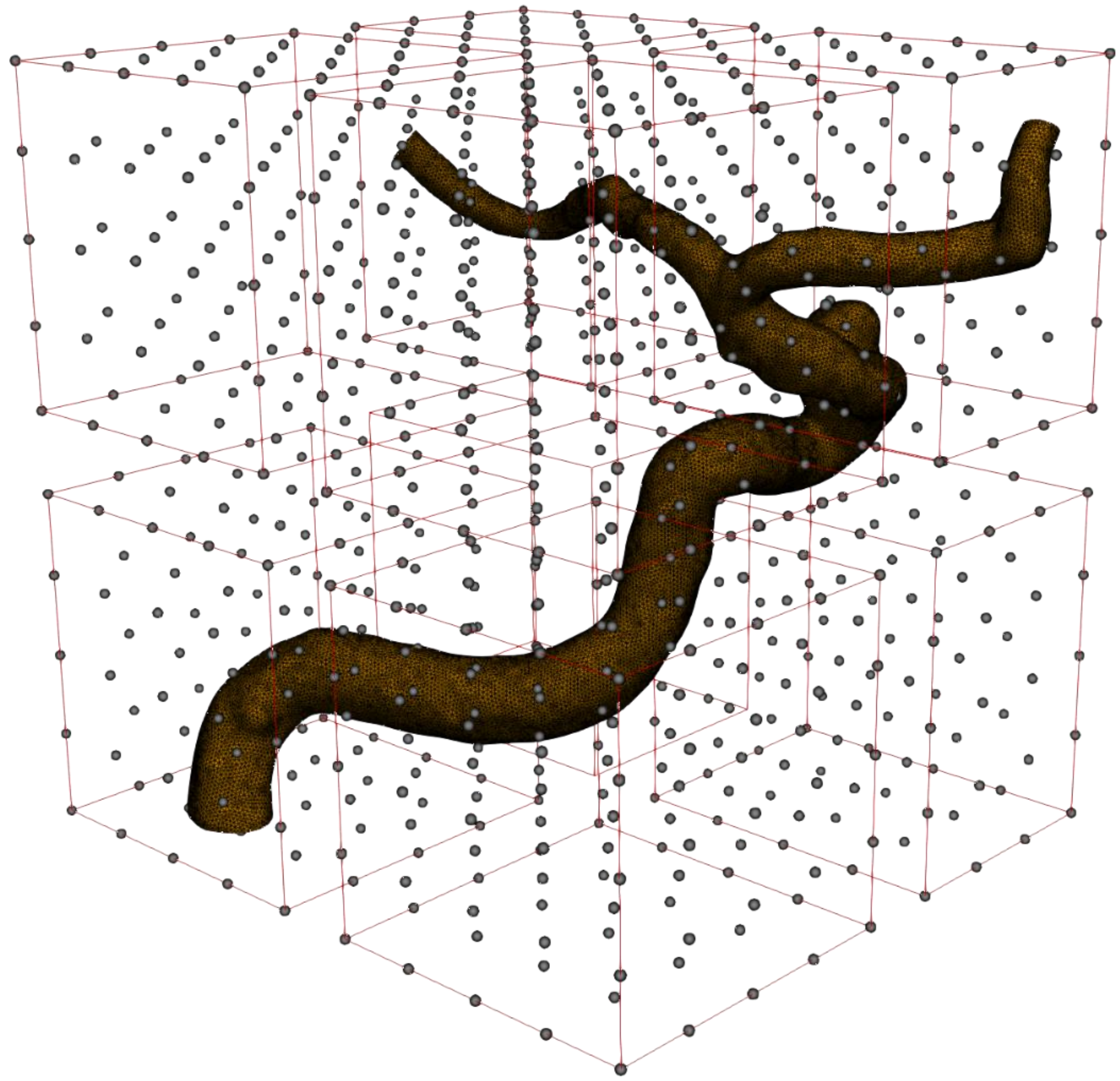

Fig. 3: Ray-casting method applied to a complex vascular geometry. Both the geometry and bounding box are partitioned among MPI processors for improved accuracy and efficiency.

## 2.3. Interpolated bounce-back scheme

One significant advantage of the LBM is its intuitive handling of the no-slip boundary condition at solid walls. In this method, maintaining this condition relies on the behavior of particle groups within the computational cells. As shown in Fig. 4, when fluid particles from group $f_1$ move toward a solid boundary, they travel halfway through the time step $\Delta t/2$ before encountering the wall. Upon impact, these particles are reflected into the fluid domain, retracing their path and returning to the cell center by the end of the full-time step $t + \Delta t$. This reversal of direction—from direction 1 to direction 2—naturally enforces the no-slip condition, ensuring that the fluid velocity at the boundary matches that of the solid wall.

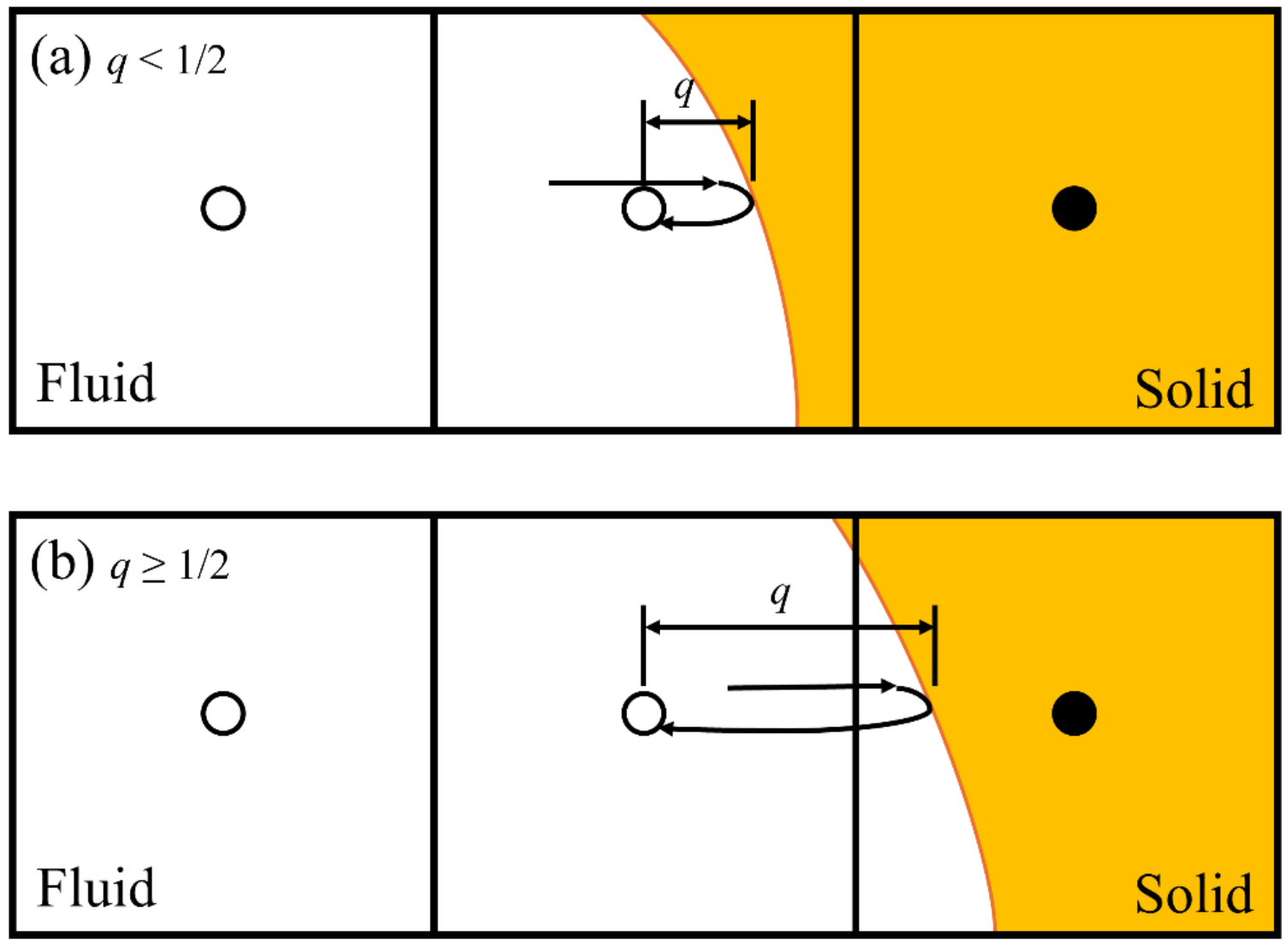

Fig. 4: Interpolated bounce-back scheme [45]

$$\tilde{f}_2(x, t + \Delta t) = \tilde{f}_1(x, t) \qquad (8)$$

This approach is known as the half-way bounce-back scheme. To enhance its applicability for complex geometries, Pan et al. [45] introduced an extended version that accommodates arbitrary curved boundaries. In cases where $q < 1/2$, an interpolation method is applied to determine the distribution function following the bounce-back process. This interpolation technique effectively improves the accuracy of the bounce-back scheme when dealing with non-planar boundary surfaces.

$$\tilde{f}_i(x, t + \Delta t) = 2q_i \tilde{f}_i(x, t) + (1 - 2q_i)\tilde{f}_i(x - c_i \Delta t, t) \qquad (9)$$

## 2.4. Inlet and outlet boundary conditions

In CFD simulations of cerebral aneurysms, defining appropriate boundary conditions at the inlet and outlet is crucial for accurately capturing physiological blood flow behavior. The inlet boundary condition typically involves prescribing a time-dependent mass flow rate or velocity profile to mimic the pulsatile nature of blood flow driven by the cardiac cycle. This ensures realistic flow patterns entering the vessel. On the other hand, the outlet boundary condition is commonly set using a pressure profile that reflects downstream vascular resistance and compliance. By incorporating these dynamic conditions, the simulation can better replicate the complex hemodynamic environment within the aneurysm and its surrounding vasculature.

Fig. 5 illustrates these conditions, with key cardiac phases — including maximum acceleration, maximum deceleration, peak systolic, and early diastolic [8] — marked to emphasize their influence on flow behavior.

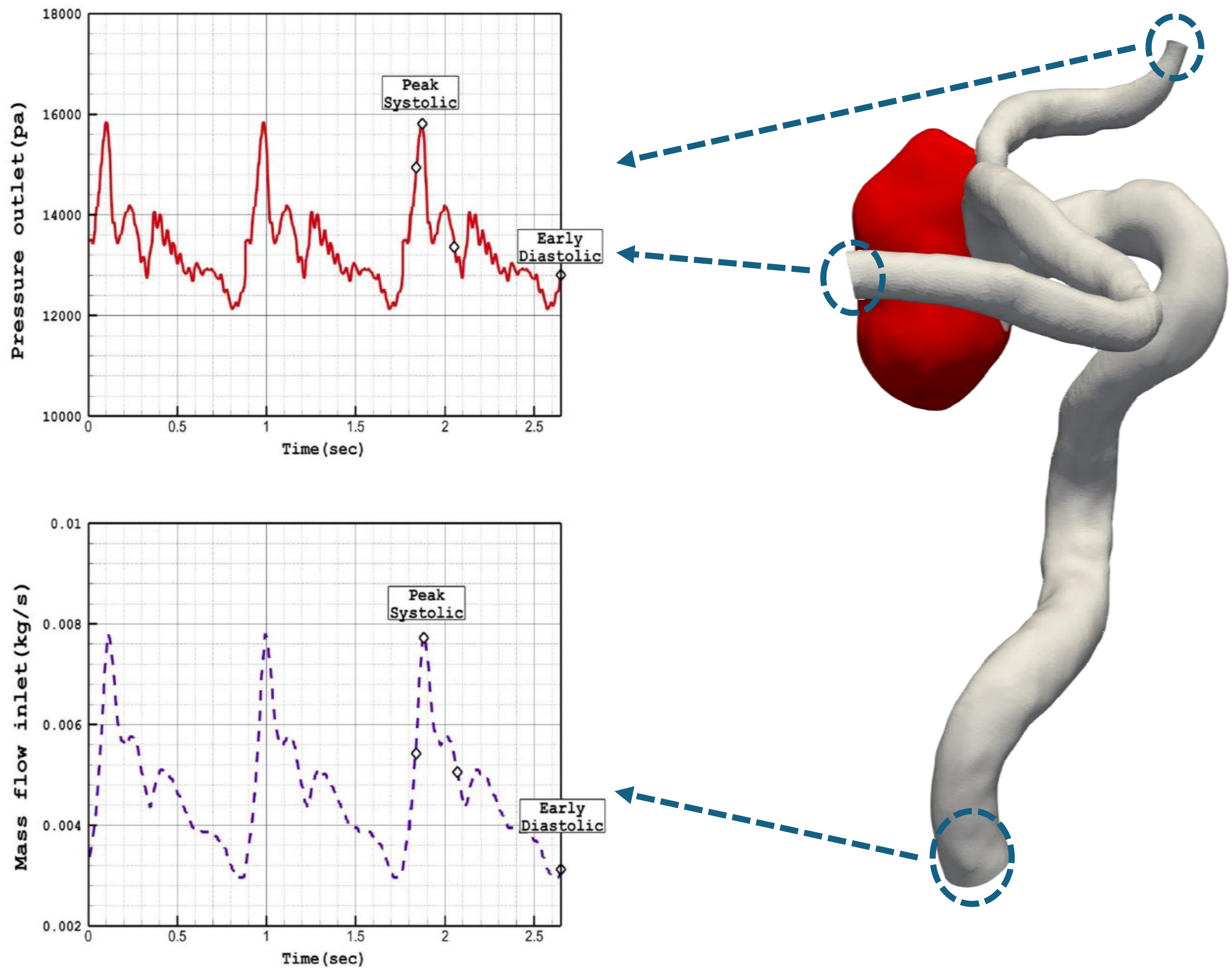

Fig. 5: Applied profile at inlet (mass flow) and outlets (pressure outlet) [8].

To set the inlet velocity and outlet pressure in our simulation, we applied the Zou-He boundary condition [46, 47]. This method defines boundary planes using a normal vector ($n$), tangential vectors $t_i = c_i - (c_i.n)n$, and a bounce-back direction $f_{-i}$ where $c_{-i} = -c_i$. The outgoing population distribution $f_{-i}$ is calculated from the incoming population $f_i$ using:

$$f_{-i} = f_i - \frac{\rho}{6} c_i.V - \frac{\rho}{3} t_i.V + \frac{1}{2}\sum_{j=1}^{19} f_j(t_i.c_j)(1 - |c_j.n|) \qquad (10)$$

This approach efficiently handles velocity and pressure boundaries in the Lattice Boltzmann Method.

## 3. Results and Discussion

In this section, we present the analysis of key hemodynamic parameters, including streamlines, velocity magnitude, WSS, TAWSS, and OSI. As mentioned earlier, our study focuses on six patient-specific vascular geometries with known aneurysm rupture statuses. The top row of the following figures corresponds to patients who responded positively — 45, 48, and 56-year-old females — representing unruptured aneurysms. Conversely, the bottom row represents patients aged 64, 66, and 84 years, all of whom experienced an aneurysm rupture. For clarity, we introduce the abbreviations **U** for unruptured and **R** for ruptured aneurysms. For example, **45U** refers to the 45-year-old patient with an unruptured aneurysm, while **64R** denotes the 64-year-old patient with a ruptured aneurysm. This notation will be used throughout the results section for consistency.

### 3.1. Mesh Independency analysis

A mesh independence study is essential in computational simulations, including LBM simulations, to ensure that the results are not influenced by the discretization of the computational domain. This process involves running simulations with varying mesh resolutions and observing if the results converge to a consistent solution. Achieving mesh independence confirms that the numerical solution accurately represents the physical problem, free from artifacts introduced by the mesh itself

To ensure both accuracy and computational efficiency, a mesh independence study was conducted on case 48U, a representative case selected for its characteristic vascular geometry. Performing this test for every patient-specific case would be impractical; therefore, this case served as a benchmark. The study examined mesh sizes of 100, 50, 40, 30, and 20 $\mu m$. The results showed that the solution stabilized at 40 $\mu m$, confirming that further mesh refinement had a negligible effect. Consequently, this resolution was applied to all other cases simulations to maintain consistency, optimize computational efficiency, and ensure accuracy in capturing flow characteristics within the cerebral vasculature.

### 3.2. Hemodynamics Parameters

Time-averaged hemodynamic parameters related to WSS are essential for understanding the complex intracranial hemodynamics that influence aneurysm behavior. In this study, we investigated WSS, TAWSS, OSI, RRT, and ECAP to evaluate their roles in aneurysm evolution.

The WSS was calculated following the standard definition, which represents the tangential force per unit area exerted by blood flow on the vessel wall [48].

$$\tau_i = \frac{\mu\omega}{C_s^2 \rho} f_\alpha^{neq} C_{\alpha j} n_j (C_{\alpha i} - C_{\alpha k} n_i j_k) \quad (11)$$

TAWSS provides the mean magnitude of the WSS vector throughout the cardiac cycle. This measure captures the overall shear stress exerted on the vessel walls over time. Here, T represents the duration of the cardiac cycle, while $\tau_w$ denotes the instantaneous wall shear stress vector at any given moment.

$$TAWSS = \frac{1}{T}\int_0^T |\tau_w| dt \qquad (12)$$

OSI is a dimensionless parameter that captures the directional variations of the WSS vector throughout the cardiac cycle relative to the primary flow direction.

$$OSI = \frac{1}{2}\left(1 - \frac{\left|\int_0^T \tau_w dt\right|}{\int_0^T |\tau_w| dt}\right) \qquad (13)$$

The Endothelial Cell Activation Potential (ECAP) establishes a correlation between TAWSS and OSI values.

$$ECAP = OSI/TAWSS \qquad (14)$$

This parameter is commonly used to describe the artery wall's susceptibility to thrombosis. Elevated ECAP values, which are linked to endothelial vulnerability, are typically observed in regions with high OSI and low TAWSS values [49, 50].

Finally, the RRT parameter is a key measure for evaluating how long blood flow remains near the vessel wall [51]. Its significance in aneurysm research lies in its ability to highlight regions where blood particles remain for extended periods. Such prolonged residence times are closely linked to an increased risk of thrombus formation, which may eventually lead to aneurysm occlusion [52].

$$RRT = \frac{1}{(1-2.OSI).TAWSS} \qquad (15)$$

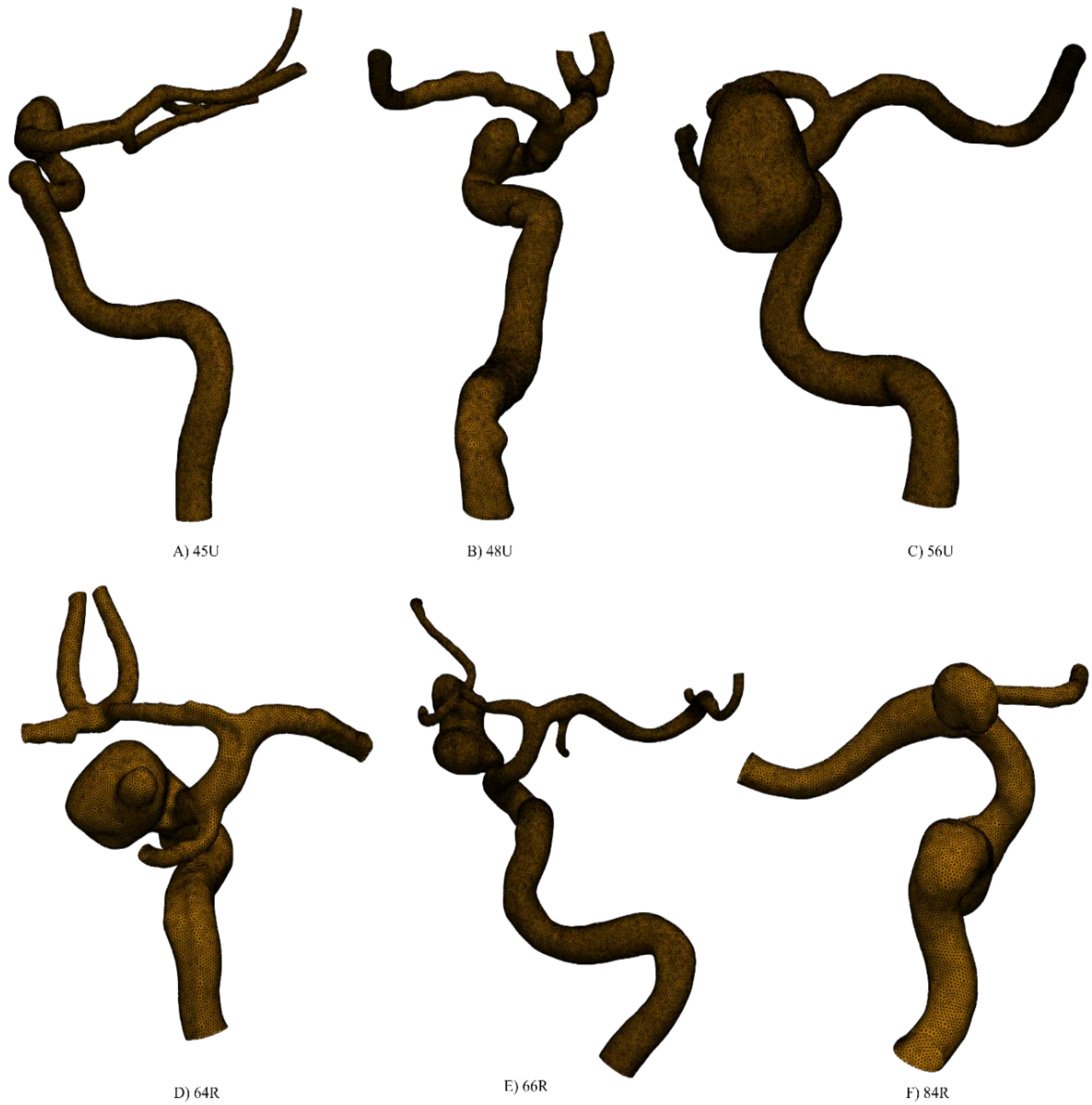

Fig. 6: 3D vascular geometries of unruptured cases (45U, 48U, 56U; top row) and ruptured cases (64R, 66R, 84R; bottom row).

Fig. 6 represents the 3D reconstructed geometries of cerebral aneurysms used for hemodynamic analysis. The models illustrate diverse aneurysm shapes, sizes, and locations within the intracranial vasculature. These geometries were generated from medical imaging data and processed to ensure smooth, accurate vessel walls for CFD simulations. The variations in aneurysm morphology reflect the complexity of aneurysmal growth and rupture risk assessment. Such reconstructed models are essential for analyzing flow dynamics, including WSS, OSI, RRT, and ECAP, to better understand the hemodynamic environment in aneurysm-prone regions.

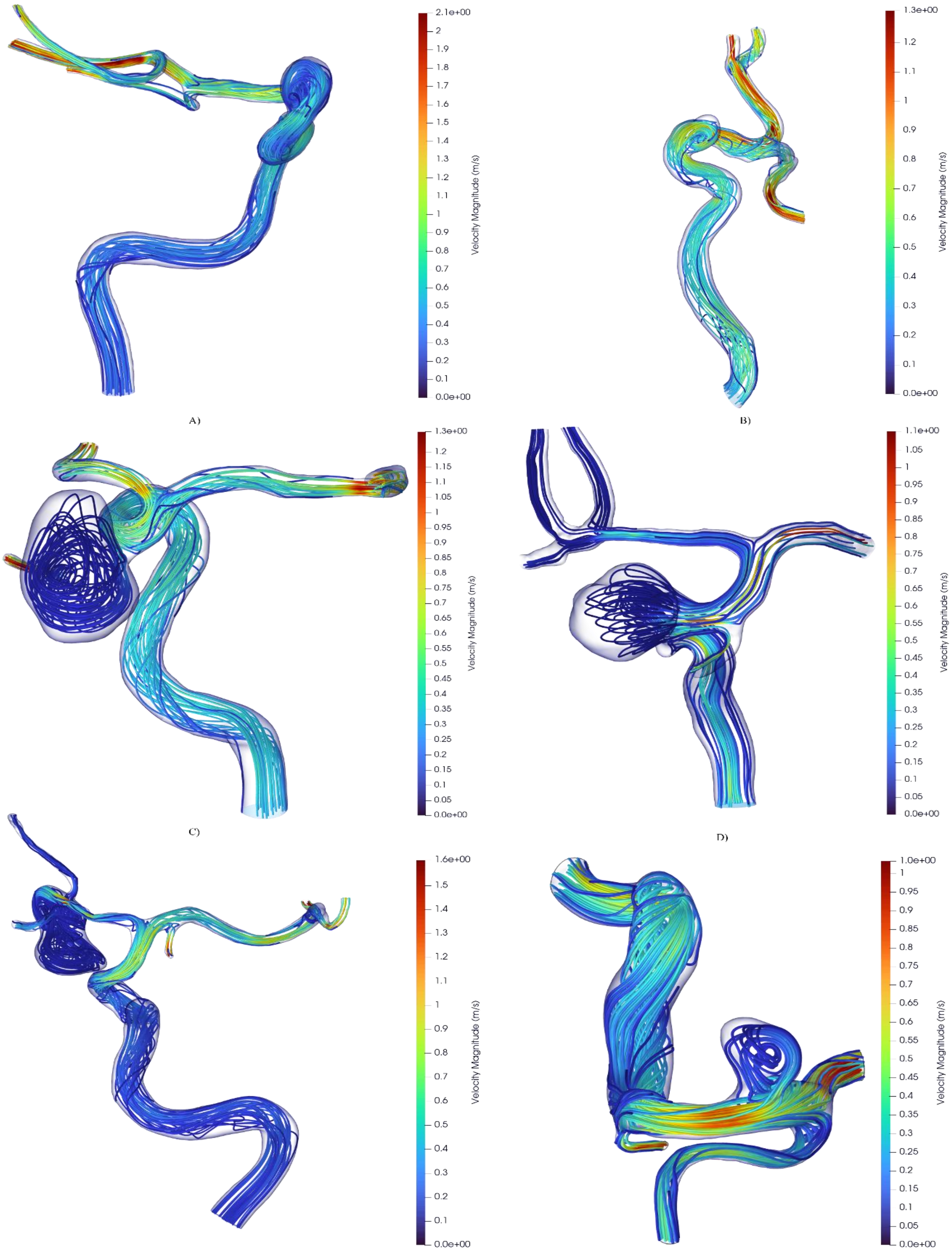

Fig. 7: Velocity Magnitude and streamline visualization for six patients, A) 45U, B) 48U, C) 56U, D) 64R, E) 66R, F) 84R

Fig. 7 shows velocity magnitude and streamlines for six patient-specific aneurysm models, including three unruptured cases (45U, 48U, 56U) and three ruptured cases (64R, 66R, 84R), to compare flow dynamics between the two groups.

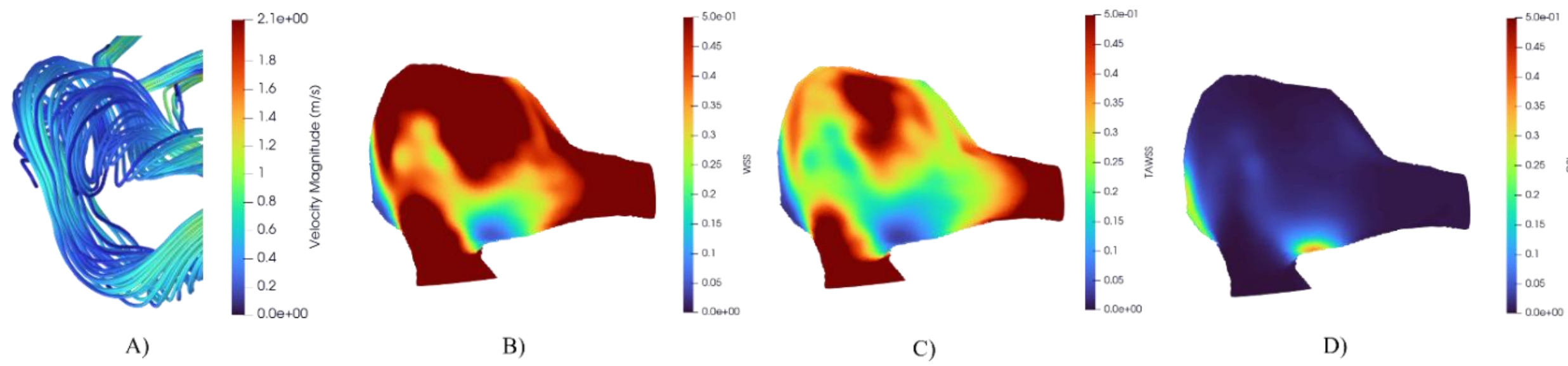

Fig. 8: Hemodynamic visualization for a 45-year-old female patient (unruptured) at systole: (A) Velocity streamlines, (B) WSS, (C) TAWSS, and (D) OSI.

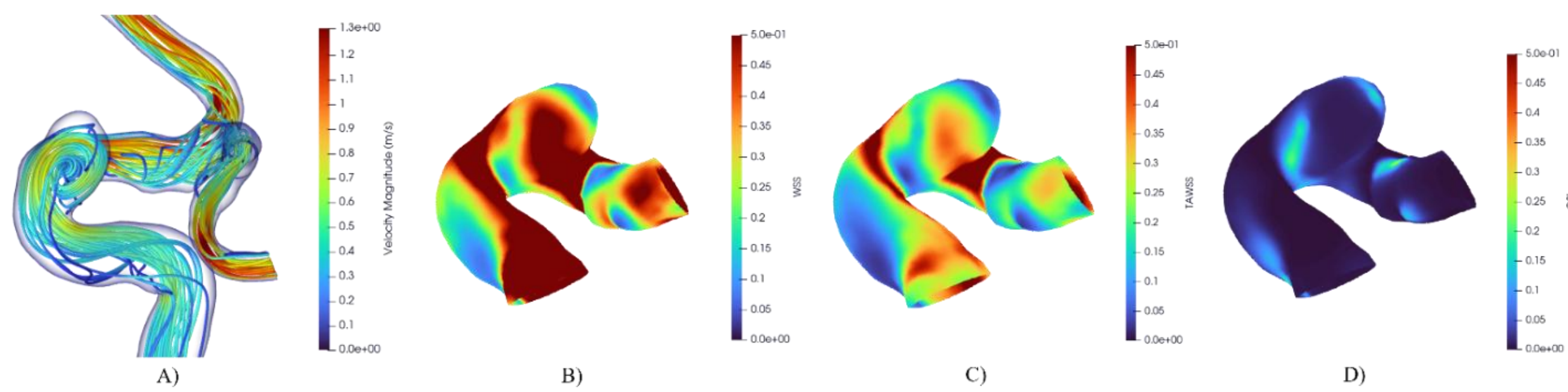

Fig. 9: Hemodynamic visualization for a 48-year-old female patient (unruptured) at systole: (A) Velocity streamlines, (B) WSS, (C) TAWSS, and (D) OSI.

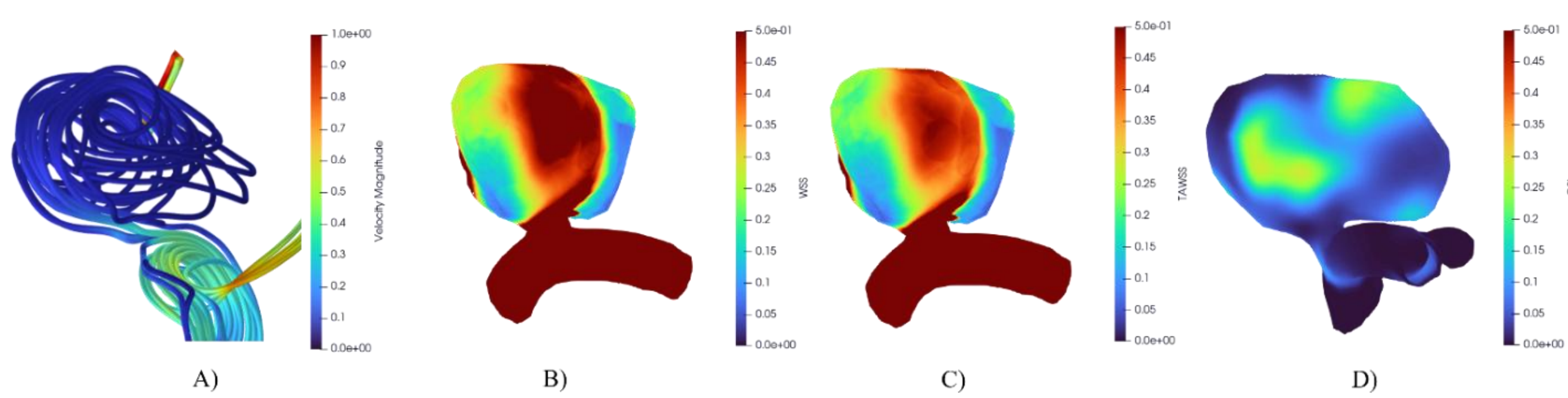

Fig. 10: Hemodynamic visualization for a 56-year-old female patient (unruptured) at systole: (A) Velocity streamlines, (B) WSS, (C) TAWSS, and (D) OSI.

Fig. 8, Fig. 9 and Fig. 10 presents hemodynamic visualizations at systole for three unruptured aneurysm cases (45U, 48U, and 56U), showing velocity streamlines, wall shear stress (WSS), time-averaged wall shear stress (TAWSS), and oscillatory shear index (OSI) to characterize flow patterns and surface-level hemodynamic metrics.

Fig. 11, Fig. 12 and Fig. 13 shows hemodynamic visualizations at systole for three ruptured aneurysm cases (64R, 66R, and 84R), including velocity streamlines, wall shear stress (WSS), time-averaged wall shear stress (TAWSS), and oscillatory shear index (OSI), to illustrate the flow dynamics and surface-level stress distributions.

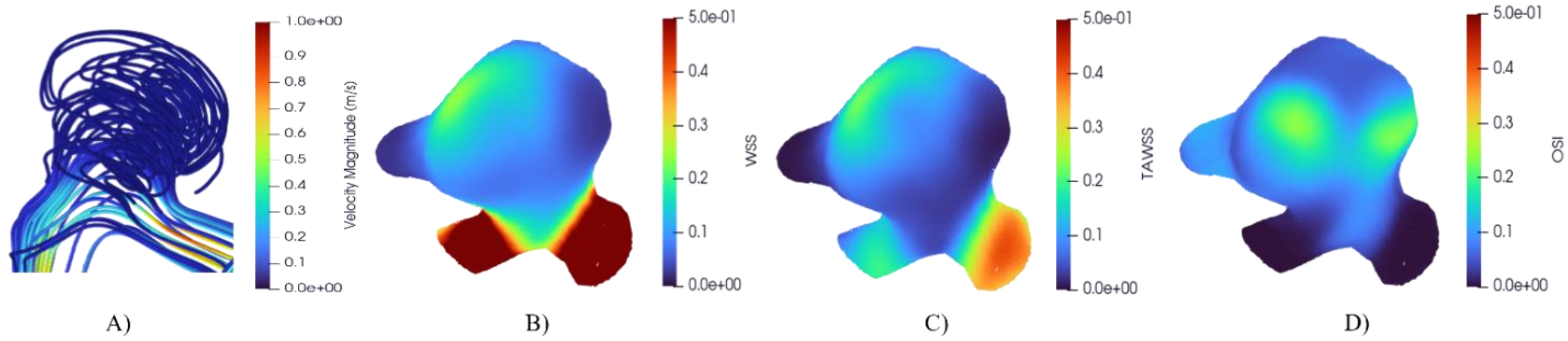

Fig. 11: Hemodynamic visualization for a 64-year-old female patient (ruptured) at systole: (A) Velocity streamlines, (B) WSS, (C) TAWSS, and (D) OSI.

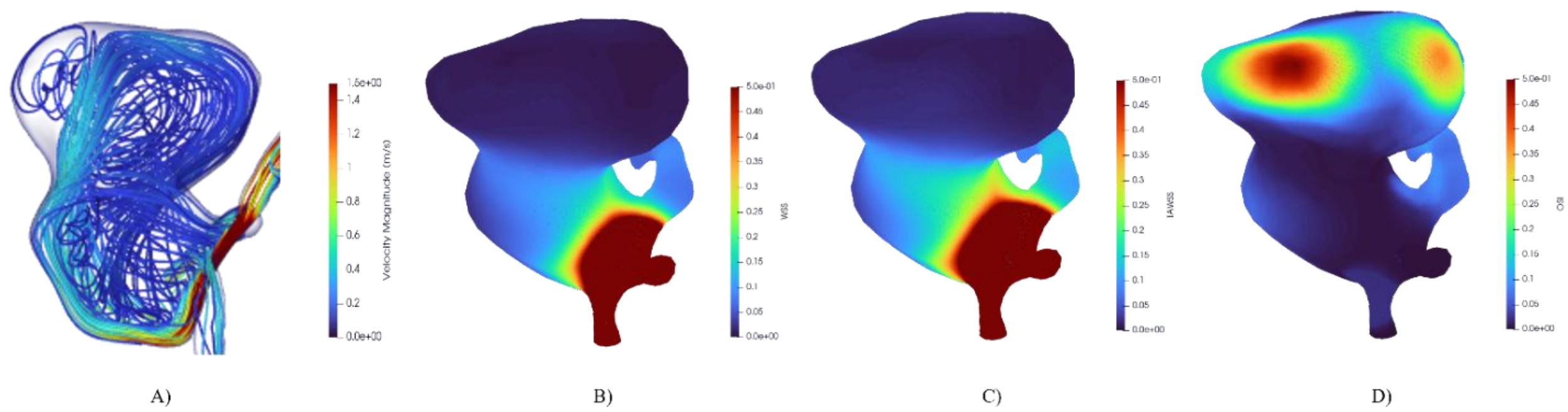

Fig. 12: Hemodynamic visualization for a 66-year-old female patient (ruptured) at systole: (A) Velocity streamlines, (B) WSS, (C) TAWSS, and (D) OSI.

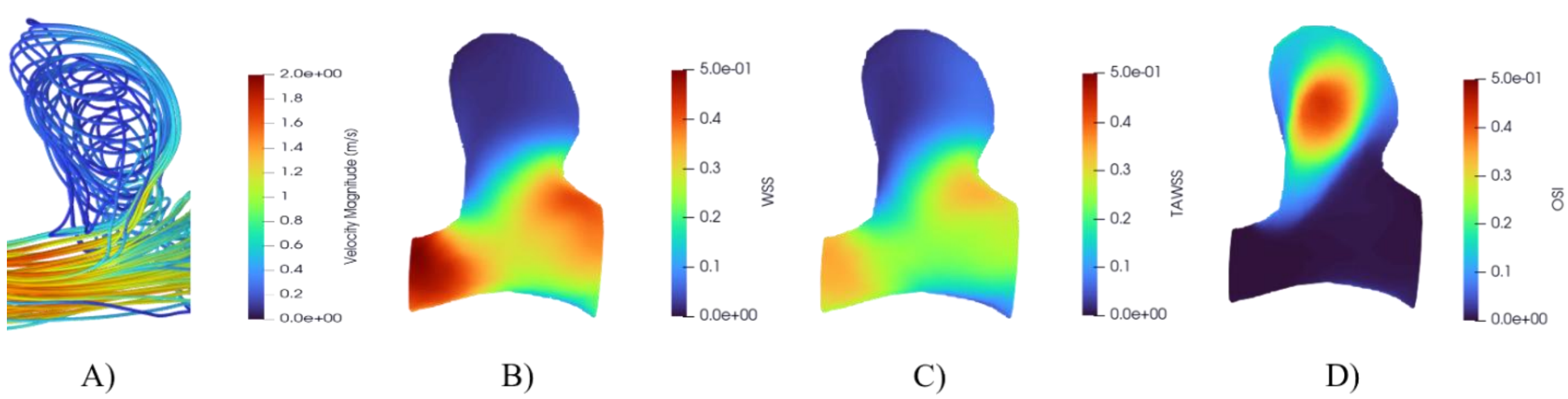

Fig. 13: Hemodynamic visualization for a 84-year-old female patient (ruptured) at systole: (A) Velocity streamlines, (B) WSS, (C) TAWSS, and (D) OSI

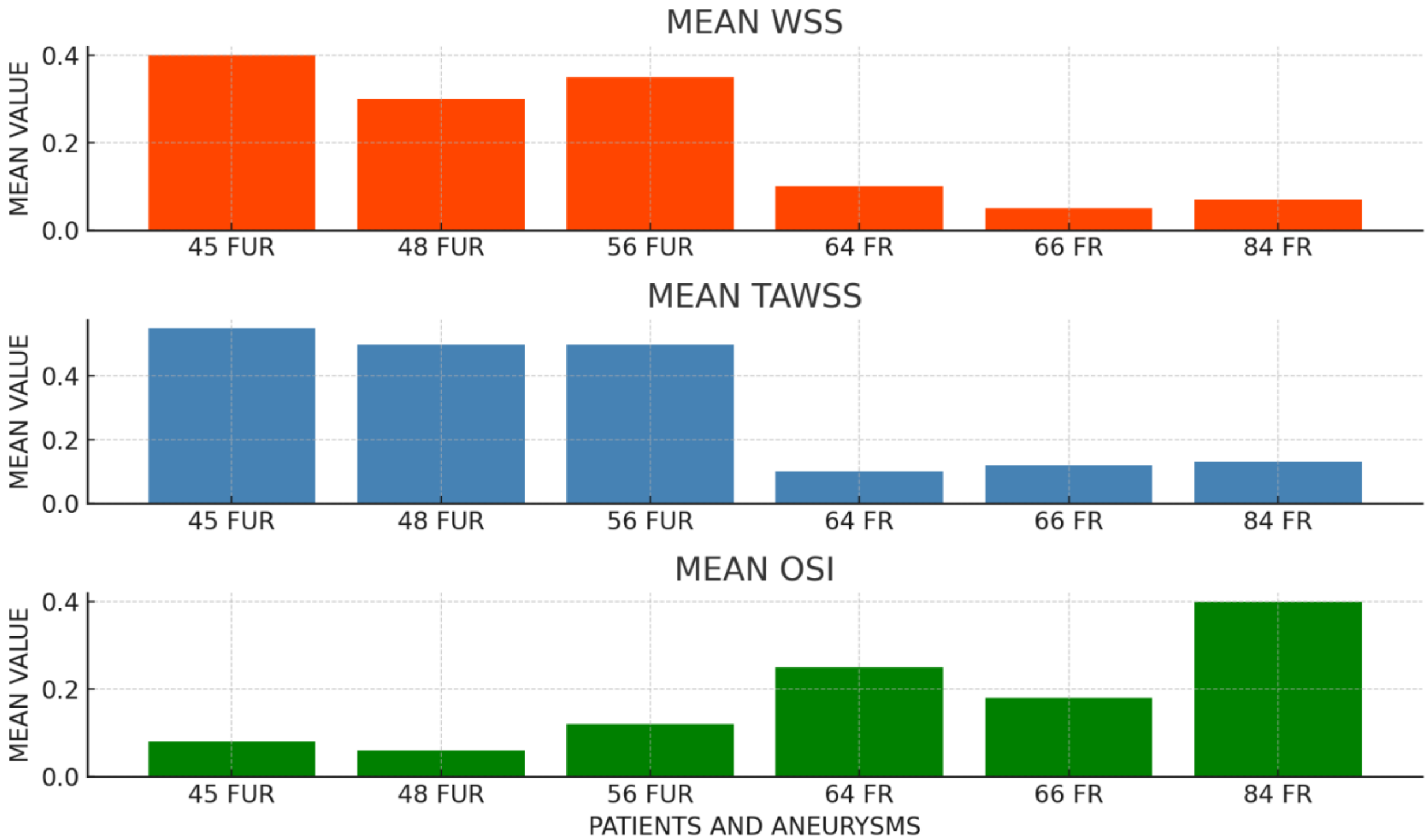

Fig. 14: Mean values of WSS, TAWSS, and OSI averaged over each aneurysm sac.

Fig. 14 presents the mean values of WSS, TAWSS, and OSI for all six aneurysm cases. The comparison highlights clear hemodynamic differences between unruptured (45 FUR, 48 FUR, 56 FUR) and ruptured (64 FR, 66 FR, 84 FR) aneurysms.

In this section, we analyzed the hemodynamic parameters of unruptured and ruptured aneurysm cases using WSS, TAWSS, and OSI. The first three cases (Fig. 8, Fig. 9 and Fig. 10) in the presented images correspond to unruptured aneurysms. These cases exhibited notably higher WSS and TAWSS values compared to the ruptured group (Fig. 11, Fig. 12 and Fig. 13). Elevated WSS values have been previously associated with aneurysm stability due to enhanced shear forces acting on the vessel walls, promoting endothelial cell alignment and vascular wall integrity [9, 20, 53-60]. Furthermore, OSI values in the unruptured cases were consistently lower. Lower OSI values are typically linked to stable flow conditions that reduce the likelihood of disturbed, oscillatory flow, which has been shown to promote aneurysm rupture [61-64]. Interestingly, the locations of higher OSI regions matched the regions with well-defined, organized vortex centers. This observation aligns with prior studies suggesting that the central region of vortices often correlates with areas of stronger, stable shear forces [62, 65-67].

In contrast, the final three cases (Fig. 11, Fig. 12 and Fig. 13) correspond to ruptured aneurysm patients. These cases demonstrated lower WSS and TAWSS values, consistent with prior findings that weakened shear forces are linked to endothelial dysfunction and potential aneurysm wall degradation [68-72]. Conversely, OSI values were substantially higher in these cases. Elevated OSI values are indicative of disturbed and oscillatory flow patterns, which are known to trigger inflammatory processes that weaken the aneurysm wall and increase rupture risk [61, 73-81].

Similar to the unruptured cases, the regions with the highest OSI values were observed to align closely with vortex centers. This suggests a potential link between unsteady flow patterns, vortex formation, and increased OSI in regions prone to rupture. Such findings are consistent with previous research highlighting the role of vortex dynamics in aneurysm instability [63, 82-84].

To further examine our findings, we demonstrated the mean values of WSS, TAWSS, and OSI on systole phase of cardiac cycle in Fig. 14. The results show that unruptured aneurysm cases (45 FUR, 48 FUR, 56 FUR) exhibit higher Mean WSS and TAWSS values, which are commonly linked to stable flow conditions that enhance vascular wall integrity. Conversely, the ruptured aneurysm cases (64 FR, 66 FR, 84 FR) present lower mean WSS and TAWSS values, indicating reduced shear forces that may contribute to endothelial dysfunction and wall weakening. Additionally, mean OSI values are notably higher in the ruptured cases, reflecting disturbed and oscillatory flow patterns known to promote aneurysm rupture risk.

In addition to WSS, TAWSS, and OSI, we further analyzed the Relative Residence Time (RRT) and Endothelial Cell Activation Potential (ECAP) to enhance our understanding of the hemodynamic environment in aneurysm cases. The results (Fig. 15 and Fig. 16) revealed distinct differences between unruptured and ruptured aneurysm groups in these parameters.

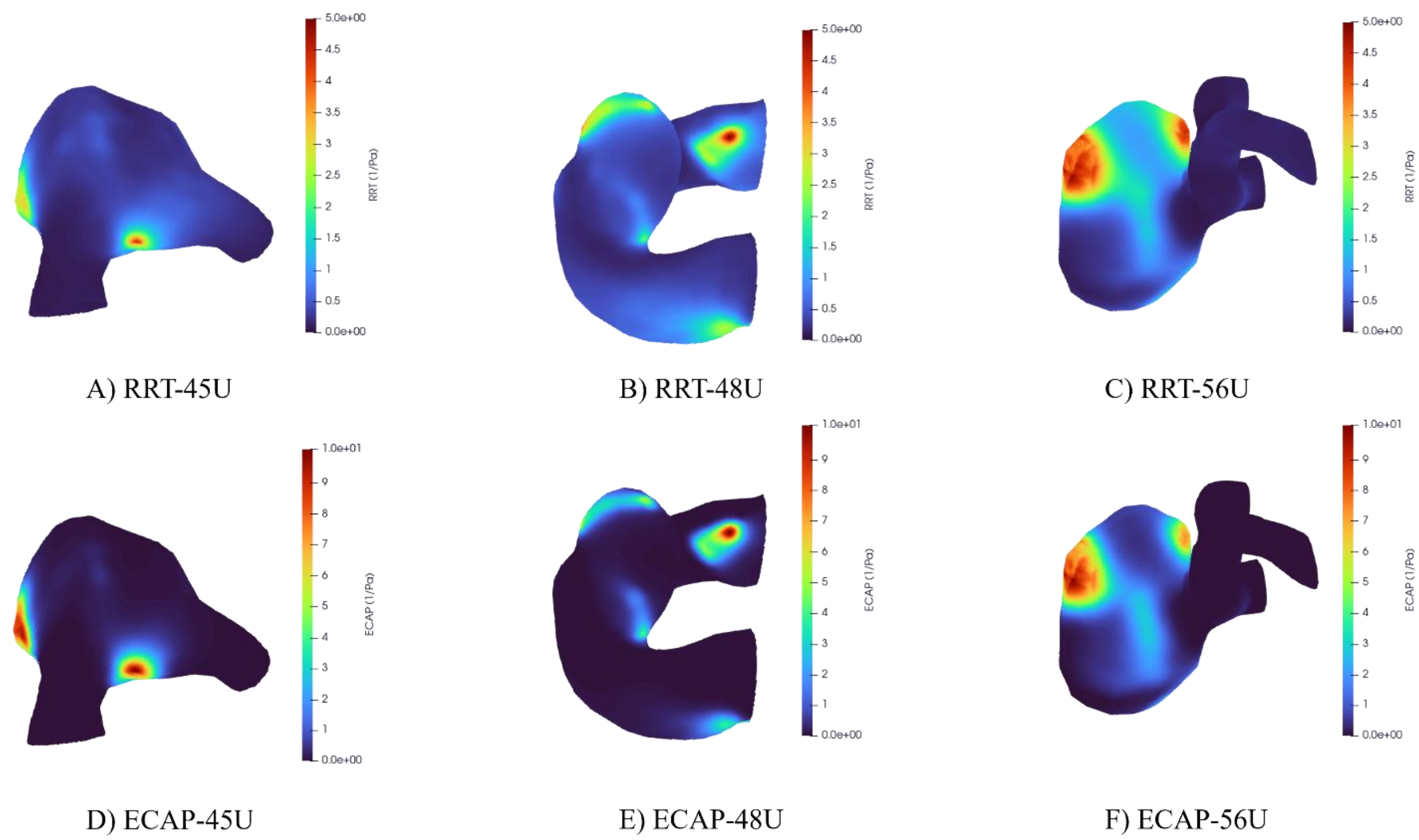

Fig. 15: RRT (top) and ECAP (bottom) distributions showing elevated values in disturbed flow regions.

Unruptured aneurysm cases exhibited lower RRT values, which aligned with previous studies suggesting that reduced residence time is indicative of more efficient blood flow patterns that mitigate prolonged exposure of the endothelial surface to potentially harmful shear stress gradients [52, 59, 85-87]. Lower RRT is generally associated with improved wall stability, as rapid blood flow reduces the likelihood of blood stagnation, platelet aggregation, and thrombus formation. Conversely, ruptured aneurysm cases showed elevated RRT values, consistent with prior findings linking prolonged residence time to disturbed flow [88-91], increased endothelial stress, and subsequent vascular remodeling that may predispose the aneurysm to rupture.

Regarding ECAP, unruptured aneurysm cases demonstrated lower ECAP values, indicating reduced endothelial activation. Lower ECAP is typically linked to stable flow conditions that limit endothelial stress, thereby promoting vascular wall integrity [83-85]. In contrast, ruptured cases exhibited elevated ECAP values, which are associated with heightened endothelial activation driven by disturbed and oscillatory flow patterns. This increase in ECAP suggests a more inflamed and vulnerable endothelial state, which has been shown to contribute to aneurysm wall degradation and rupture risk [49, 50, 85].

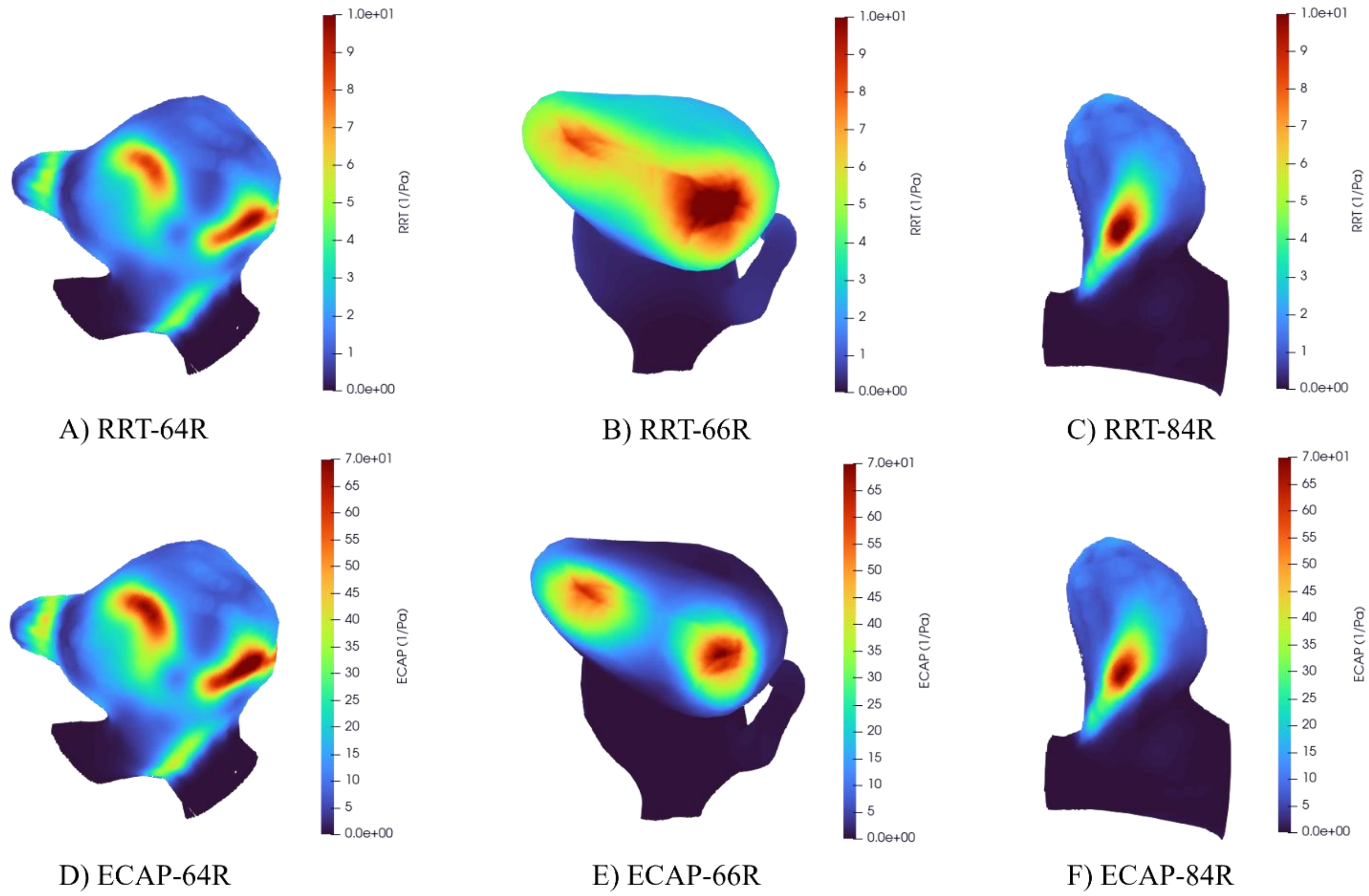

Fig. 16: RRT (top) and ECAP (bottom) distributions showing elevated values in disturbed flow regions.

Interestingly, in both unruptured and ruptured aneurysm groups, the regions with the highest RRT values tended to align with areas of increased OSI and organized vortex centers. This observation supports the hypothesis that disturbed flow patterns characterized by prolonged residence time and oscillatory shear stress contribute to aneurysm instability and rupture risk. Similarly, regions with elevated ECAP were often correlated with these unstable flow zones, reinforcing the notion that endothelial activation plays a crucial role in aneurysm pathophysiology.

These combined findings suggest that RRT and ECAP, alongside WSS, TAWSS, and OSI, provide valuable insights into the complex hemodynamic factors influencing aneurysm stability and rupture risk. Integrating these parameters offers a more comprehensive understanding of aneurysm pathophysiology, potentially aiding in improved risk assessment and clinical decision-making.

## Conclusion

This study highlights the critical role of hemodynamic parameters in assessing cerebral aneurysm stability and rupture risk. Our analysis demonstrates that unruptured aneurysms are characterized by higher WSS and TAWSS values, alongside lower OSI, RRT, and ECAP values—features indicative of stable flow conditions that promote endothelial health and vascular wall integrity. Conversely, ruptured aneurysm cases exhibit reduced WSS and TAWSS values, coupled with elevated OSI, RRT, and ECAP values, suggesting disturbed, oscillatory flow patterns that contribute to endothelial dysfunction and wall weakening.

Notably, the alignment of high OSI and RRT regions with vortex centers underscores the role of localized disturbed flow in aneurysm instability. The elevated ECAP values observed in ruptured cases further emphasize the link between endothelial activation and aneurysm wall degradation, reinforcing the importance of endothelial response as a potential marker for rupture risk.

These findings suggest that a comprehensive evaluation of WSS, TAWSS, OSI, RRT, and ECAP provides a more robust framework for understanding aneurysm pathophysiology. Integrating these parameters may improve rupture risk assessment and enhance clinical strategies for aneurysm monitoring and treatment. Future research should explore the combined predictive value of these hemodynamic markers across larger patient cohorts to refine risk stratification models and guide intervention strategies.

**Conflict of Interest**

The authors declare that they have no conflict of interest.

**Ethical Approval**

This study did not involve any experiments with human participants or animals performed by the authors. All patient-specific vascular geometries used in this research were obtained from publicly available repositories and were fully anonymized prior to acquisition. Therefore, ethical approval and informed consent were not required.

**Funding**

No external funding was received for this study.

**Competing Interests**

The authors declare that they have no competing interests.

## References

1. Burlakoti, A., et al., *Trend of cerebral aneurysms over the past two centuries: need for early screening*. BMJ open, 2024. **14**(2): p. e081290.
2. Khasawneh, M., B. Eby, and A. Vellimana, *Duplicated Middle Cerebral Artery: Prevalence and Clinical Significance in Aneurysm and Stroke Management*. Stroke: Vascular and Interventional Neurology, 2024. **4**: p. e12984_247.
3. Wei, J., et al., *Knowledge-augmented deep learning for segmenting and detecting cerebral aneurysms with CT angiography: a multicenter study*. Radiology, 2024. **312**(2): p. e233197.
4. Navaei, R.L. and H. Tehrani, *An Integrated Lab on a CD Microfluidic Platform for High-Efficiency Blood Cell Separation and Passive Mixing*. arXiv preprint arXiv:2507.01977, 2025.
5. Tang, Y., et al., *Transition of intracranial aneurysmal wall enhancement from high to low wall shear stress mediation with size increase: A hemodynamic study based on 7T magnetic resonance imaging*. Heliyon, 2024. **10**(9).
6. Shimogonya, Y., S. Fukuda, and C.A.S. Group, *Role of disturbed wall shear stress in the development of cerebral aneurysms*. Journal of Biomechanics, 2024. **176**: p. 112355.
7. Futami, K., et al., *Characterization of Maximum Wall Shear Stress Points in Unruptured Cerebral Aneurysms Using Four-dimensional Flow Magnetic Resonance Imaging*. Clinical Neuroradiology, 2024. **34**(4): p. 899-906.
8. Huo, H. and Y. Chang, *Hemodynamic study of the ICA aneurysm evolution to attain the cerebral aneurysm rupture risk*. Scientific Reports, 2024. **14**(1): p. 8984.
9. Shen, F., et al., *Influence of neck width on transient flow characteristics in saccular intracranial aneurysm models*. Acta Mechanica Sinica, 2025. **41**(4): p. 324196.
10. Brindise, M., et al. *The influence of transient physiological factors on patient-specific cerebral aneurysm hemodynamics*. in *APS Division of Fluid Dynamics Meeting Abstracts*. 2024.
11. Abdelghafar, A., et al., *Comparison between ruptured anterior choroidal artery aneurysms and ruptured intracranial aneurysms in other locations in relation to aneurysm dimensions at rupture*. Acta Neurochirurgica, 2025. **167**(1): p. 12.
12. Sanchez, S., et al., *Morphological characteristics of ruptured brain aneurysms: a systematic literature review and meta-analysis*. Stroke: Vascular and Interventional Neurology, 2023. **3**(2): p. e000707.
13. Cui, Y., et al., *Aneurysm morphological prediction of intracranial aneurysm rupture in elderly patients using four-dimensional CT angiography*. Clinical Neurology and Neurosurgery, 2021. **208**: p. 106877.
14. Jelen, M.B., et al., *Psychological and functional impact of a small unruptured intracranial aneurysm diagnosis: a mixed-methods evaluation of the patient journey*. Stroke: Vascular and Interventional Neurology, 2023. **3**(1): p. e000531.

15. Sabernaeemi, A., et al., *Influence of stent-induced vessel deformation on hemodynamic feature of bloodstream inside ICA aneurysms*. Biomechanics and Modeling in Mechanobiology, 2023. **22**(4): p. 1193-1207.
16. Zhang, C., et al., *Hematoma evacuation via image-guided para-corticospinal tract approach in patients with spontaneous intracerebral hemorrhage*. Neurology and Therapy, 2021. **10**: p. 1001-1013.
17. Vu, D.L., et al., *Hemodynamic Characteristics in Ruptured and Unruptured Intracranial Aneurysms: A Prospective Cohort Study Utilizing the AneurysmFlow Tool*. American Journal of Neuroradiology, 2025.
18. Dong, L., et al., *Age-related impairment of structure and function of iliac artery endothelium in rats is improved by elevated fluid shear stress*. Medical science monitor: international medical journal of experimental and clinical research, 2019. **25**: p. 5127.
19. Jiang, H., et al., *The influence of sac centreline on saccular aneurysm rupture: Computational study*. Scientific Reports, 2023. **13**(1): p. 11288.
20. Lampropoulos, D.S. and M. Hadjinicolaou, *Investigating Hemodynamics in Intracranial Aneurysms with Irregular Morphologies: A Multiphase CFD Approach*. Mathematics, 2025. **13**(3): p. 505.
21. Rüttgers, M., et al., *Patient-specific lattice-Boltzmann simulations with inflow conditions from magnetic resonance velocimetry measurements for analyzing cerebral aneurysms*. Computers in Biology and Medicine, 2025. **187**: p. 109794.
22. Xu, G., et al., *Application of Computational Fluid Dynamic Simulation of Parent Blood Flow in the Embolization of Unruptured A1 Aneurysms*. World Neurosurgery, 2025. **193**: p. 696-705.
23. Rahma, A.G. and T. Abdelhamid, *Hemodynamic and fluid flow analysis of a cerebral aneurysm: a CFD simulation*. SN Applied Sciences, 2023. **5**(2): p. 62.
24. Jin, Z.-H., et al., *CFD investigations of the blood hemodynamic inside internal cerebral aneurysm (ICA) in the existence of coiling embolism*. Alexandria Engineering Journal, 2023. **66**: p. 797-809.
25. Fillingham, P., et al., *Improving the accuracy of computational fluid dynamics simulations of coiled cerebral aneurysms using finite element modeling*. Journal of biomechanics, 2023. **157**: p. 111733.
26. Huang, A. and W. Zhou, *Mn-based cGAS-STING activation for tumor therapy*. Chinese Journal of Cancer Research, 2023. **35**(1): p. 19.
27. Mao, X., et al., *Tissue resident memory T cells are enriched and dysfunctional in effusion of patients with malignant tumor*. Journal of Cancer, 2023. **14**(7): p. 1223.
28. Fung, Y.-c., *Biomechanics: mechanical properties of living tissues*. 2013: Springer Science & Business Media.
29. Zhou, G., et al., *Association of wall shear stress with intracranial aneurysm rupture: systematic review and meta-analysis*. Scientific reports, 2017. **7**(1): p. 5331.
30. Zhang, Y., et al., *Low wall shear stress is associated with the rupture of intracranial aneurysm with known rupture point: case report and literature review*. BMC neurology, 2016. **16**: p. 1-4.

31. Miura, Y., et al., *Low wall shear stress is independently associated with the rupture status of middle cerebral artery aneurysms.* Stroke, 2013. **44**(2): p. 519-521.
32. Kulcsár, Z., et al., *Hemodynamics of cerebral aneurysm initiation: the role of wall shear stress and spatial wall shear stress gradient.* American Journal of neuroradiology, 2011. **32**(3): p. 587-594.
33. Che, Y., et al., *Rupture prediction of medium to large-sized abdominal aortic aneurysm combining wall shear stress-related parameters and anatomical characteristics: A computational, experimental, and statistical analysis.* Physics of Fluids, 2025. **37**(1).
34. Morel, S., et al., *Effects of low and high aneurysmal wall shear stress on endothelial cell behavior: differences and similarities.* Frontiers in Physiology, 2021. **12**: p. 727338.
35. Veeturi, S.S., et al., *Hemodynamic analysis shows high wall shear stress is associated with intraoperatively observed thin wall regions of intracranial aneurysms.* Journal of Cardiovascular Development and Disease, 2022. **9**(12): p. 424.
36. Kumar, V.S. and V.S. Kumar, *High Wall Shear incites cerebral aneurysm formation and low Wall Shear stress propagates cerebral aneurysm growth.* Journal of Neurology Research, 2023. **13**(1): p. 1-11.
37. Cho, K.-C., *The current limitations and advanced analysis of hemodynamic study of cerebral aneurysms.* Neurointervention, 2023. **18**(2): p. 107-113.
38. Shen, X.-Y., et al., *Numerical simulation of blood flow effects on rupture of aneurysm in middle cerebral artery.* International Journal of Modern Physics C, 2022. **33**(03): p. 2250030.
39. Kuzmin, A. and J. Derksen. *Introduction to the lattice Boltzmann method*. in *LBM Workshop*. 2011.
40. Bhatnagar, P.L., E.P. Gross, and M. Krook, *A model for collision processes in gases. I. Small amplitude processes in charged and neutral one-component systems.* Physical review, 1954. **94**(3): p. 511.
41. Hagspiel, K.D., et al., *Computed tomography angiography and magnetic resonance angiography imaging of the mesenteric vasculature.* Techniques in vascular and interventional radiology, 2015. **18**(1): p. 2-13.
42. Hartung, M.P., T.M. Grist, and C.J. François, *Magnetic resonance angiography: current status and future directions.* Journal of Cardiovascular Magnetic Resonance, 2011. **13**(1): p. 19.
43. Khaniki, M.A.L., et al., *Vision transformer with feature calibration and selective cross-attention for brain tumor classification.* Iran Journal of Computer Science, 2024: p. 1-13.
44. Kim, W.Y., et al., *Coronary magnetic resonance angiography for the detection of coronary stenoses.* New England Journal of Medicine, 2001. **345**(26): p. 1863-1869.
45. Pan, C., L.-S. Luo, and C.T. Miller, *An evaluation of lattice Boltzmann schemes for porous medium flow simulation.* Computers & fluids, 2006. **35**(8-9): p. 898-909.
46. Hecht, M. and J. Harting, *Implementation of on-site velocity boundary conditions for D3Q19 lattice Boltzmannsimulations.* Journal of Statistical Mechanics: Theory and Experiment, 2010. **2010**(01): p. P01018.

47. Zou, Q. and X. He, *On pressure and velocity boundary conditions for the lattice Boltzmann BGK model.* Physics of fluids, 1997. **9**(6): p. 1591-1598.
48. Matyka, M., Z. Koza, and Ł. Mirosław, *Wall orientation and shear stress in the lattice Boltzmann model.* Computers & Fluids, 2013. **73**: p. 115-123.
49. Boniforti, M.A., et al., *Image-based numerical investigation in an impending abdominal aneurysm rupture.* Fluids, 2022. **7**(8): p. 269.
50. Di Achille, P., et al., *A haemodynamic predictor of intraluminal thrombus formation in abdominal aortic aneurysms.* Proceedings of the Royal Society A: Mathematical, Physical and Engineering Sciences, 2014. **470**(2172): p. 20140163.
51. Himburg, H.A., et al., *Spatial comparison between wall shear stress measures and porcine arterial endothelial permeability.* American Journal of Physiology-Heart and Circulatory Physiology, 2004. **286**(5): p. H1916-H1922.
52. Rayz, V., et al., *Flow residence time and regions of intraluminal thrombus deposition in intracranial aneurysms.* Annals of biomedical engineering, 2010. **38**: p. 3058-3069.
53. Baek, H., et al., *Flow instability and wall shear stress variation in intracranial aneurysms.* Journal of the Royal Society Interface, 2010. **7**(47): p. 967-988.
54. Brinjikji, W., et al., *Hemodynamic differences between unstable and stable unruptured aneurysms independent of size and location: a pilot study.* Journal of neurointerventional surgery, 2017. **9**(4): p. 376-380.
55. Meng, H., et al., *High WSS or low WSS? Complex interactions of hemodynamics with intracranial aneurysm initiation, growth, and rupture: toward a unifying hypothesis.* American Journal of Neuroradiology, 2014. **35**(7): p. 1254-1262.
56. Morel, S., P. Bijlenga, and B.R. Kwak, *Intracranial aneurysm wall (in) stability–current state of knowledge and clinical perspectives.* Neurosurgical review, 2022. **45**(2): p. 1233-1253.
57. Sanchez, S., et al., *Comprehensive morphomechanical and wall enhancement analysis of intracranial aneurysms.* European Radiology, 2025: p. 1-10.
58. Sforza, D.M., et al., *Hemodynamics in growing and stable cerebral aneurysms.* Journal of neurointerventional surgery, 2016. **8**(4): p. 407-412.
59. Sheikh, M.A.A., A.S. Shuib, and M.H.H. Mohyi, *A review of hemodynamic parameters in cerebral aneurysm.* Interdisciplinary Neurosurgery, 2020. **22**: p. 100716.
60. Xiang, J., et al., *Hemodynamic–morphological discriminant models for intracranial aneurysm rupture remain stable with increasing sample size.* Journal of neurointerventional surgery, 2016. **8**(1): p. 104-110.
61. Aggarwal, S., et al., *Abdominal aortic aneurysm: A comprehensive review.* Experimental & Clinical Cardiology, 2011. **16**(1): p. 11.
62. Chatziprodromou, I., et al., *Haemodynamics and wall remodelling of a growing cerebral aneurysm: a computational model.* Journal of biomechanics, 2007. **40**(2): p. 412-426.
63. Iaccarino, G., et al., *Ischemic neoangiogenesis enhanced by β2-adrenergic receptor overexpression: a novel role for the endothelial adrenergic system.* Circulation research, 2005. **97**(11): p. 1182-1189.
64. Savastano, L.E., et al., *Biology of cerebral aneurysm formation, growth, and rupture*, in *Intracranial aneurysms*. 2018, Elsevier. p. 17-32.

65. HJ, S., *Pathophysiology of development and rupture of cerebral aneurysms*. Acta Neurochir Suppl, 1990.
66. Steiger, H.J., et al., *Strength, elasticity and viscoelastic properties of cerebral aneurysms*. Heart and vessels, 1989. **5**: p. 41-46.
67. Vlak, M.H., et al., *Independent risk factors for intracranial aneurysms and their joint effect: a case-control study*. Stroke, 2013. **44**(4): p. 984-987.
68. Chang, H.S., *Simulation of the natural history of cerebral aneurysms based on data from the International Study of Unruptured Intracranial Aneurysms*. Journal of neurosurgery, 2006. **104**(2): p. 188-194.
69. Chatziprodromou, I., D. Poulikakos, and Y. Ventikos, *On the influence of variation in haemodynamic conditions on the generation and growth of cerebral aneurysms and atherogenesis: a computational model*. Journal of biomechanics, 2007. **40**(16): p. 3626-3640.
70. Sforza, D.M., C.M. Putman, and J.R. Cebral, *Hemodynamics of cerebral aneurysms*. Annual review of fluid mechanics, 2009. **41**(1): p. 91-107.
71. Sharma, K.V., R. Straka, and F.W. Tavares, *Current status of Lattice Boltzmann Methods applied to aerodynamic, aeroacoustic, and thermal flows*. Progress in Aerospace Sciences, 2020. **115**: p. 100616.
72. Wiebers, D.O., et al. *Pathogenesis, natural history, and treatment of unruptured intracranial aneurysms*. in *Mayo clinic proceedings*. 2004. Elsevier.
73. Broderick, J.P., et al., *Initial and recurrent bleeding are the major causes of death following subarachnoid hemorrhage*. Stroke, 1994. **25**(7): p. 1342-1347.
74. Gaetani, P.G., et al., *Metalloproteases and intracranial vascular lesions*. Neurological research, 1999. **21**(4): p. 385-390.
75. Greving, J.P., et al., *Development of the PHASES score for prediction of risk of rupture of intracranial aneurysms: a pooled analysis of six prospective cohort studies*. The Lancet Neurology, 2014. **13**(1): p. 59-66.
76. Inoue, T., et al., *Annual rupture risk of growing unruptured cerebral aneurysms detected by magnetic resonance angiography*. Journal of neurosurgery, 2012. **117**(1): p. 20-25.
77. Juvela, S., M. Porras, and O. Heiskanen, *Natural history of unruptured intracranial aneurysms: a long-term follow-up study*. Journal of neurosurgery, 1993. **79**(2): p. 174-182.
78. Kleinloog, R., et al., *Risk factors for intracranial aneurysm rupture: a systematic review*. Neurosurgery, 2018. **82**(4): p. 431-440.
79. MacDonald, D.J., H.M. Finlay, and P.B. Canham, *Directional wall strength in saccular brain aneurysms from polarized light microscopy*. Annals of biomedical engineering, 2000. **28**: p. 533-542.
80. Mitchell, P. and J. Jakubowski, *Estimate of the maximum time interval between formation of cerebral aneurysm and rupture*. Journal of Neurology, Neurosurgery & Psychiatry, 2000. **69**(6): p. 760-767.
81. Zhang, X., et al., *Cerebral microbleeds could be independently associated with intracranial aneurysm rupture: a cross-sectional population-based study*. World neurosurgery, 2018. **115**: p. e218-e225.

82. AJ, M., *Cerebral Aneurysm Multicenter European Onyx (CAMEO) trial: results of a prospective observational study in 20 European centers.* AJNR Am J Neuroradiol, 2004. **25**: p. 39-51.
83. Byrne, J.V., et al., *Early experience in the treatment of intra-cranial aneurysms by endovascular flow diversion: a multicentre prospective study.* PloS one, 2010. **5**(9): p. e12492.
84. Lylyk, P., et al., *Curative endovascular reconstruction of cerebral aneurysms with the pipeline embolization device: the Buenos Aires experience.* Neurosurgery, 2009. **64**(4): p. 632-643.
85. Boniforti, M.A., G. Vittucci, and R. Magini, *Endovascular Treatment of Intracranial Aneurysm: The Importance of the Rheological Model in Blood Flow Simulations.* Bioengineering, 2024. **11**(6): p. 522.
86. Sugiyama, S.-i., et al., *Relative residence time prolongation in intracranial aneurysms: a possible association with atherosclerosis.* Neurosurgery, 2013. **73**(5): p. 767-776.
87. Reza, M.M.S. and A. Arzani, *A critical comparison of different residence time measures in aneurysms.* Journal of biomechanics, 2019. **88**: p. 122-129.
88. Xiang, J., et al., *Hemodynamic–morphologic discriminants for intracranial aneurysm rupture.* Stroke, 2011. **42**(1): p. 144-152.
89. Zhu, Y., et al., *Assessing the risk of intracranial aneurysm rupture using computational fluid dynamics: a pilot study.* Frontiers in Neurology, 2023. **14**: p. 1277278.
90. Uchikawa, H., et al., *Aneurysmal Inflow Rate Coefficient Predicts Ultra-early Rebleeding in Ruptured Intracranial Aneurysms: Preliminary Report of a Computational Fluid Dynamics Study.* Neurologia medico-chirurgica, 2023. **63**(10): p. 450-456.
91. Xiang, J., et al., *CFD: computational fluid dynamics or confounding factor dissemination? The role of hemodynamics in intracranial aneurysm rupture risk assessment.* American Journal of Neuroradiology, 2014. **35**(10): p. 1849-1857.